\documentclass{article}
\usepackage{spconf,amsmath,graphicx}

\newcommand{\bi}[1]{\ensuremath{\boldsymbol{#1}}}   

\usepackage{amssymb}
\usepackage{enumerate}
\usepackage{graphicx}
\usepackage{multirow}
\usepackage{arydshln}
\usepackage{color}
\usepackage{float}
\usepackage{bm}
\usepackage{comment}
\usepackage{cite}
\usepackage{amsmath}
\usepackage{algorithmic}
\usepackage{array}

\newlength\savedwidth
\newcommand{\wcline}[1]{\noalign{\global\savedwidth\arrayrulewidth\global\arrayrulewidth 1.0pt} \cline{#1}
\noalign{\global\arrayrulewidth\savedwidth}}

\title{Sound Event Detection Using Graph Laplacian Regularization\\Based on Event Co-occurrence}
\name{Keisuke Imoto$^{\dagger}$ and Seisuke Kyochi$^{\ddagger}$}
\address{$^{\dagger}$ Ritsumeikan University, Japan, $^{\ddagger}$ The University of Kitakyushu, Japan}
\begin{document}
%
\maketitle
\begin{abstract}
The types of sound events that occur in a situation are limited, and some sound events are likely to co-occur; for instance, ``dishes'' and ``glass jingling.''
In this paper, we propose a technique of sound event detection utilizing graph Laplacian regularization taking the sound event co-occurrence into account.
In the proposed method, sound event occurrences are represented as a graph whose nodes indicate the frequency of event occurrence and whose edges indicate the co-occurrence of sound events.
This graph representation is then utilized for sound event modeling, which is optimized under an objective function with a regularization term considering the graph structure.
Experimental results obtained using TUT Sound Events 2016 development, 2017 development, and TUT Acoustic Scenes 2016 development indicate that the proposed method improves the detection performance of sound events by 7.9 percentage points compared to that of the conventional CNN-BiGRU-based method in terms of the segment-based F1-score.
Moreover, the results show that the proposed method can detect co-occurring sound events more accurately than the conventional method.
\end{abstract}
\begin{keywords}
Sound event detection, graph Laplacian regularization, sound event co-occurrence, convolutional recurrent neural network, acoustic scene
\end{keywords}
%
%
\vspace{-8pt}
\section{Introduction}
\label{sec:intro}
\vspace{-8pt}
Sound event detection (SED), in which the onsets and offsets of sound events are detected and the types of sounds are identified \cite{Imoto_AST2018_01}, has significant potential for use in many applications such as monitoring elderly people or infants \cite{Peng_ICME2009_01,Guyot_ICASSP2013_01}, automatic surveillance \cite{Radhakrishnan_WASPAA2005_01,Harma_ICME2005_01,Ntalampiras_ICASSP2009_01}, and media retrieval \cite{Jin_INTERSPEECH2012_01}.

SED typically falls into two categories: monophonic and polyphonic SED.
In monophonic SED, it is assumed that multiple sound events do not occur simultaneously; thus, a monophonic SED system only detects at most one sound event in each time section.
However, in a real-life situation, since multiple sound events tend to overlap in time, a monophonic SED system has limited performance in a real-life environment.
To overcome this limitation, many polyphonic SED systems, which can detect multiple overlapping sound events, have been developed.

One approach to polyphonic SED is to use non-negative matrix factorization (NMF) \cite{Dessein_MIG2013_01,Komatsu_DCASE2016_01}.
In the SED approach based on NMF, a polyphonic sound is decomposed into a product of a basis and activation matrices, where each basis vector and activation vector respectively indicate a single sound event and the active duration of the corresponding sound event.
SED systems based on neural networks have also been developed \cite{Hershey_ICASSP2017_01,Cakir_TASLP2017_01,Hayashi_TASLP2017_01,Kothinti_DCASE2018_01}.
For instance, a convolutional neural network (CNN) is widely used for SED \cite{Hershey_ICASSP2017_01}. 
More recently, many methods using a recurrent neural network (RNN) or convolutional recurrent neural network (CRNN), which can capture temporal information of sound events, have been developed \cite{Cakir_TASLP2017_01,Hayashi_TASLP2017_01,Kothinti_DCASE2018_01}.
These methods enable successful analysis of overlapping sound events with reasonable performance.
However, when the number of types of sound events to be analyzed increases, these approaches require a large training dataset.

\begin{figure*}[t]
\centering
\begin{tabular}{c}
\hspace{-16pt}
\begin{minipage}{0.615\hsize}
\centering
\includegraphics[width=0.96\columnwidth]{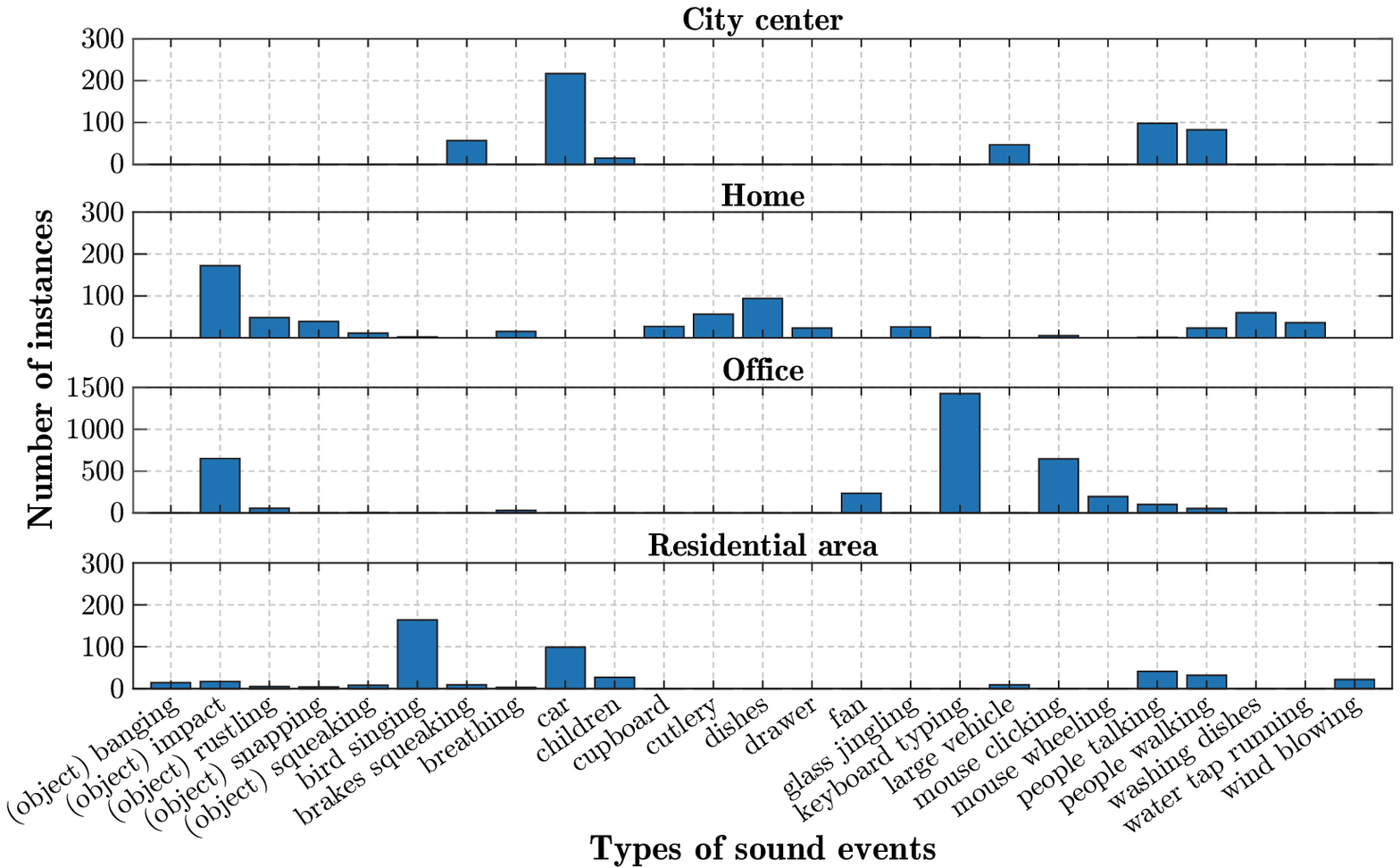}
\vspace{-6pt}
\caption{Histogram of sound event instances for each acoustic scene}
\label{fig:numofinstance}
\end{minipage}
\hspace{6pt}
\begin{minipage}{0.37\hsize}
\centering
\includegraphics[width=1.02\columnwidth]{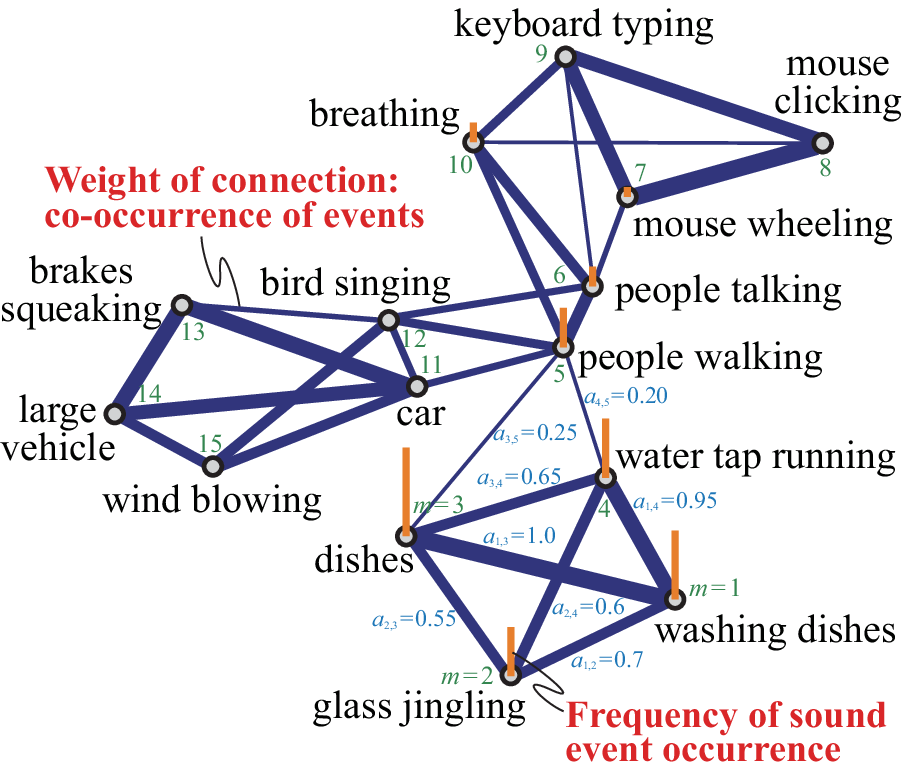}
\vspace{-12pt}
\caption{Concept of graph representation of sound event occurrences}
\label{fig:incidence}
\end{minipage}
\end{tabular}
\end{figure*}

On the other hand, as shown in Fig.~\ref{fig:numofinstance}, the number of types of sound events occurring in a single situation (acoustic scene) is limited and some sound events co-occur.
For instance, the sound events ``dishes'' and ``glass jingling'' tend to co-occur, and ``car'' and ``brakes squeaking'' are also likely to co-occur.
By considering this in the model training of sound events, we expect to be able to model sound events efficiently and effectively with limited sound data \cite{Mesaros_EUSIPCO2011_01,Imoto_TASLP2019_01}.
However, conventional methods cannot be integrated into the state-of-the-art neural network-based method.
Thus, in this paper, we propose an SED approach based on a neural network that considers the co-occurrence of sound events in each sound clip.
To consider the co-occurrence of sound events, we introduce graph Laplacian regularization into the objective function of a neural network.

The rest of this paper is organized as follows.
In section 2, a conventional SED approach based on a CRNN is introduced.
In section 3, the proposed approach to SED, in which the co-occurrence of sound events can be considered, is discussed.
In section 4, we report experiments conducted to evaluate the performance of SED by the proposed and conventional methods,
and in section 5, we summarized and conclude this paper.
%
%
%
\vspace{-8pt}
\section{Conventional Sound Event Detection Based on Recurrent Neural Networks}
\label{sec:crnn}
\vspace{-8pt}
In this section, we review conventional SED approaches based on neural networks.
For polyphonic SED, CNN architectures are often used \cite{Hershey_ICASSP2017_01}.
In CNN-based SED, the time-frequency representation of the acoustic feature ${\bf V} \in \mathbb{R}^{D \times T}$ is fed to a convolutional layer, where $D$ and $T$ are the dimension of the acoustic feature and the number of time frames of the input feature, respectively.
This layer convolutes the input feature map with two-dimensional filters; then, max pooling is conducted to reduce the dimension of the feature map.
The CNN architecture allows robust feature extraction against time and frequency shifts, which often occur in SED.

To model time correlations explicitly, an RNN has been applied to SED in some works \cite{Cakir_TASLP2017_01,Hayashi_TASLP2017_01,Kothinti_DCASE2018_01}.
In particular, it has been reported that neural networks combining a CNN and bidirectional gated recurrent unit (BiGRU) successfully detected sound events.
In the CNN-BiGRU-based approaches, the acoustic feature ${\bf V}$ is also fed to the convolutional layer.
The output of the convolutional layer ${\bf X} \in \mathbb{R}^{D' \times T \times C}$ is then concatenated as ${\bf X}_{concat} = ({\bf x}_{1}, {\bf x}_{2}, \ldots{\bf x}_{t}, \ldots, {\bf x}_{T}) \in \mathbb{R}^{(D' \cdot C)\times T}$, and then ${\bf X}_{concat}$ is fed to the BiGRU layer, where $C$ is the number of filters of the convolution layer.
In the BiGRU layer, the output vector ${\bf h}_{t}$ is calculated using the following equations:
\vspace{4pt}
\begin{align}
{\bf g}^{f}_{t} &= \sigma({\bf W}^{f}_{g} {\bf x}_{t} + {\bf U}^{f}_{g} {\bf h}_{t-1} + {\bf b}^{f}_{g}),\\[4pt]
{\bf r}^{f}_{t} &= \sigma({\bf W}^{f}_{r} {\bf x}_{t} + {\bf U}^{f}_{r} {\bf h}_{t-1} + {\bf b}^{f}_{r}),\\[4pt]
{\bf h}^{f}_{t} &= (1-{\bf g}^{f}_{t}) \odot {\bf h}_{t-1} \nonumber\\[0pt]
&\hspace{10pt}+ {\bf g}^{f}_{t} \odot \tanh ({\bf W}^{f}_{h} {\bf x}_{t} + {\bf U}^{f}_{h} ( {\bf r}^{f}_{h} {\bf h}_{t-1}) + {\bf b}^{f}_{h}),\\[4pt]
{\bf g}^{b}_{t} &= \sigma({\bf W}^{b}_{g} {\bf x}_{t} + {\bf U}^{b}_{g} {\bf h}_{t+1} + {\bf b}^{b}_{g}),\\[4pt]
{\bf r}^{b}_{t} &= \sigma({\bf W}^{b}_{r} {\bf x}_{t} + {\bf U}^{b}_{r} {\bf h}_{t+1} + {\bf b}^{b}_{r}),\\[4pt]
{\bf h}^{b}_{t} &= (1-{\bf g}^{b}_{t}) \odot {\bf h}_{t+1} \nonumber\\[0pt]
&\hspace{10pt}+ {\bf g}^{b}_{t} \odot \tanh ({\bf W}^{b}_{h} {\bf x}_{t} + {\bf U}^{b}_{h} ( {\bf r}^{b}_{h} {\bf h}_{t+1}) + {\bf b}^{b}_{h}),
\vspace{-19pt}
\end{align}
\vspace{-18pt}
\begin{align}
{\bf h}_{t} &=
\begin{bmatrix}
{\bf h}^{f}_{t}\\[3pt]
{\bf h}^{b}_{t}
\end{bmatrix},
\label{eq:bigru}
\end{align}
\vspace{-2pt}
where superscripts $f$ and $b$ indicate forward and backward networks, respectively.
Subscripts $t$, $g$, and $r$ indicate the time index, update gate, and reset gate, respectively.
${\bf g}$, ${\bf r}$, $\odot$, and $\sigma$ indicate the update gate vector, reset gate vector, Hadamard product, and sigmoid function, respectively.
${\bf W}$, ${\bf U}$, and ${\bf b}$ are parameter matrices and a bias vector.
The BiGRU layer is followed by a fully connected layer, which is the output layer of the network.
The final output of the network is calculated as
\vspace{-4pt}
\begin{align}
{\bf y}_{t} &= \sigma({\bf h}_{t}).
\label{eq:outputlayer}
\end{align}
%

\vspace{-2pt}
The CNN-BiGRU network is optimized under the following sigmoid cross-entropy objective function $E({\bi \Theta})$ using the backpropagation through time (BPTT):
\vspace{-5pt}
\begin{align}
E({\bi \Theta}) = - \sum^{T}_{t=1} {\big \{} {\bf z}_{t} \log ( {\bf y}_{t} ) + (1-{\bf z}_{t}) \log (1-{\bf y}_{t}) {\big \}},
\label{eq:crnn_objective}
\vspace{-9pt}
\end{align}
\noindent where ${\bf z}_{t}$ is a target vector of the output that indicates whether sound events are active or nonactive in time frame $t$.
%
%
\section{Sound Event Detection with Event-\\Co-occurrence-Based Regularization}
\label{sec:proposed}
\subsection{Motivation}
\label{subsec:motivation}
Conventional CRNN-based approaches achieve reasonable performances in SED when a sufficient amount of training sound data is prepared.
However, since recording and annotating environmental sounds are very time-consuming tasks \cite{Imoto_AST2018_01}, in many situations, the conventional method tends to exhibit degradation in the event detection performance.
To overcome this problem, we propose a new method using graph Laplacian regularization for SED.

As shown in Fig.~\ref{fig:numofinstance}, the number of types of sound events occurring in a single situation (acoustic scene) is limited and some sound events co-occur.
For instance, the sound events ``dishes'' and ``glass jingling'' tend to co-occur, and ``car'' and ``brakes squeaking'' are also likely to co-occur.
By considering the sound event co-occurrence in the model parameter estimation of a neural network, it is expected that sound events can be efficiently and effectively modeled with limited sound data.
%
\subsection{Sound event detection using graph Laplacian regularization}
\label{subsec:SEDwithGLR}
To consider the co-occurrence of sound events, we introduce the graph representation of sound event occurrences and a graph-based regularization technique for the modeling of sound events.

Suppose that a graph representation ${\bf G}$ of a sound event occurrence has nodes ${\bf v} \in \mathbb{R}^{M}$ and an adjacency matrix ${\bf A} \in \mathbb{R}^{M \times M}$, as shown in Fig.~\ref{fig:incidence}.
Here, $M$ is the number of types of sound events.
The weights of the nodes on the graph are the frequencies of sound event occurrences, and the weights of the edges indicate how often two sound events co-occur.
Then, the graph Laplacian matrix ${\bf L}$ \cite{Shuman_SPM2013_01} is defined
as 
\vspace{2pt}
\begin{align}
{\bf L} = {\bf \Delta} - {\bf A},
\end{align}
%
where ${\bf \Delta}$ is a diagonal, so-called degree matrix, whose diagonal elements are defined as $[\delta]_{ii} = \sum_j A_{i,j}$.

If two sound events tend to co-occur, when the two nodes corresponding to the sound events are connected with a large weight, the frequency of the sound event occurrence should have a small difference.
Thus, adding the following penalty term to the cost function of the optimization problem enables us to learn a sound event model in which we can consider the sound event co-occurrence \cite{Cai_TPAMI2011_01,Ichita_APSIPA2018_01}.
%
\begin{align}
\frac{1}{2} \sum_{i, j =0}^{M} A_{i,j} \| v_{i} - v_{j} \|^{2}\nonumber \\[-3pt]
&\hspace{-64pt}= \sum_{i = 0}^{M} v_{i} v_{i} \Delta_{i,i} - \sum_{i,j = 0}^{M} v_{i} v_{j} A_{i,j}\nonumber \\[3pt]
&\hspace{-64pt}= \mathrm{Tr} ({\bf v}^{\mathsf T} {\bf \Delta} {\bf v}) - \mathrm{Tr} ({\bf v}^{\mathsf T} {\bf A} {\bf v}) = \mathrm{Tr} ({\bf v}^{\mathsf T} {\bf L} {\bf v})
\label{eq:term1}
\end{align}
%
By integrating Eq. (\ref{eq:term1}) into Eq. (\ref{eq:crnn_objective}), we obtain the following objective function:
%
\begin{align}
E({\bi \Theta}) &= - \sum^{T}_{t=1} {\big \{} {\bf z}_{t} \log ( {\bf y}_{t} ) + (1-{\bf z}_{t}) \log (1-{\bf y}_{t}) {\big \}}\nonumber\\[-3pt]
&\hspace{40pt} + \alpha \mathrm{Tr}({\bf v}^{\mathsf T} {\bf L} {\bf v}).
\end{align}
%
By approximating the frequencies of sound event occurrences ${\bf v}$ by $\sum_{t} {\bf y}_{t}$, the objective function is finally given as
%
\begin{align}
E({\bi \Theta}) &= - \sum^{T}_{t=1} {\big \{} {\bf z}_{t} \log ( {\bf y}_{t} ) + (1-{\bf z}_{t}) \log (1-{\bf y}_{t}) {\big \}}\nonumber\\[-3pt]
&\hspace{40pt} + \alpha \mathrm{Tr}{\Big \{} {\big (} \sum_{t=1}^{T} {\bf y}_{t} {\big )}^{\! \mathsf T} {\bf L} {\big (} \sum_{t=1}^{T} {\bf y}_{t} {\big )} {\Big \}},
\end{align}
%
\noindent where $\alpha$ is the regularization weight.
Thus, we can detect appropriate sound events ${\bf y}_{t}$ while considering the co-occurrence of sound events, even when limited training data can be used for model training.
\begin{table}[t!]
\caption{Experimental conditions}
\label{tab:Condition}
\vspace{4pt}
\small
\centering
\renewcommand{\arraystretch}{1.0}
\begin{tabular}{lll}
\wcline{1-3}\\[-12pt]
\multicolumn{2}{l}{\!\!}&\vspace{-8pt} \\
\multicolumn{2}{l}{\!\!Acoustic feature}&Log mel-band energy\!\!\\
\multicolumn{2}{l}{\!\!\# dims. of acoustic feature}&64\!\!\\
\multicolumn{2}{l}{\!\!Frame length / shift}&40 ms / 20 ms\!\!\\
\multicolumn{2}{l}{\!\!Length of sequence}&500 (10 s)\!\!\\
\multicolumn{2}{l}{\!\!Regularization weight $\alpha$}&$1.0 \times 10^{-5}$\!\!\\
\cline{1-3}\\[-9pt]
\multicolumn{2}{l}{\!\!Network structure of CNN-BiGRU}&3 conv. \& 1 BiGRU layers\!\!\\
\multicolumn{2}{l}{\!\!Filter size in CNN layers}&3 $\times$ 3\!\!\\
\multicolumn{2}{l}{\!\!Pooling in CNN layers}&3 $\times$ 1 max pooling\!\!\\
\multicolumn{2}{l}{\!\!Activation function}&ReLU\!\!\\
\multicolumn{2}{l}{\!\!\# channels of CNN layers}&128, 128, 128\!\!\\
\multicolumn{2}{l}{\!\!\# GRU units}&32\!\!\\
\multicolumn{2}{l}{\!\!\# epochs for training}&150\!\!\\
\multicolumn{2}{l}{\!\!Optimizer}&Adam\!\!\\
\multicolumn{2}{l}{\!\!Thresholding}&Adaptive thresholding \cite{Xu_DCASE2017_01}\!\!\\[2pt]
\wcline{1-3}
\end{tabular}
\end{table}
\begin{table*}[t!]
\caption{Detection performance of sound events in segment-based metrics}
\label{tab:Result1}
\vspace{4pt}
\small
\centering
\renewcommand{\arraystretch}{1.0}
\begin{tabular}{lrrrrrrrrrr}
\wcline{1-11}\\[-13pt]
\!&&&&&&\vspace{-7pt} \\
\multicolumn{1}{c}{\multirow{2}{*}{\bf Method}}&\multicolumn{2}{c}{\bf Fold 1}&\multicolumn{2}{c}{\bf Fold 2}&\multicolumn{2}{c}{\bf Fold 3}&\multicolumn{2}{c}{\bf Fold 4}&\multicolumn{2}{c}{\bf Average}\\
\cline{2-11}\\[-9pt]
\!&F1 score\!\!&\!\!Error rate&F1-score\!\!&\!\!Error rate&F1-score\!\!&\!\!Error rate&F1-score\!\!&\!\!Error rate&F1-score\!\!&\!\!Error rate\\
\wcline{1-11}\\[-9pt]
\!CNN&48.67\%\!\!&\!\!0.708&31.36\%\!\!&\!\!0.829&33.11\%\!\!&\!\!0.813&23.55\%\!\!&\!\!0.899&34.17\%\!\!&\!\!0.812\\
\!CNN-GRU&51.00\%\!\!&\!\!0.672&36.64\%\!\!&\!\!0.795&35.95\%\!\!&\!\!0.797&34.70\%\!\!&\!\!0.864&39.57\%\!\!&\!\!0.782\\
\!CNN-BiGRU&53.10\%\!\!&\!\!0.652&35.10\%\!\!&\!\!0.807&38.34\%\!\!&\!\!0.769&38.42\%\!\!&\!\!{\bf 0.814}&41.24\%\!\!&\!\!0.761\\
\!CNN-BiGRU w/ GLR\!\!\!&{\bf 55.59}\%\!\!&\!\!{\bf 0.631}&{\bf 48.28\%}\!\!&\!\!{\bf 0.742}&{\bf 50.39\%}\!\!&\!\!{\bf 0.678}&{\bf 42.39\%}\!\!&\!\!0.820&{\bf 49.16}\%\!\!&\!\!{\bf 0.718}\\
\wcline{1-11}
\end{tabular}
\vspace{-4pt}
\end{table*}
%
%
\vspace{-3pt}
\section{Experiments}
\label{sec:experiments}
\vspace{-5pt}
\subsection{Experimental conditions}
\label{subsec:conditions}
\vspace{-1pt}
To evaluate the performance of the proposed method, we conducted experiments with conventional neural-network-based methods and the proposed method.
For the experiments, we constructed a sound event dataset composed of part of the TUT Sound Events 2016 development, 2017 development, and TUT Acoustic Scenes 2016 development \cite{Mesaros_EUSIPCO2016_01,Mesaros_DCASE2017_01}.
From the three datasets, we extracted sound clips including four acoustic scenes, home, residential area (TUT Sound Events 2016), city center (TUT Sound Events 2017), and office (TUT Acoustic Scenes 2016), with a total duration of 192 min. of audio.
The experimental data include the 25 types of sound events listed in Fig.~\ref{fig:numofinstance}.
In this regard, because the original TUT Acoustic Scenes 2016 development datasets do not have sound event annotations for the sound clips recorded in the office environment, we annotated them using the same protocol as in \cite{Mesaros_EUSIPCO2016_01} and \cite{Mesaros_DCASE2017_01}.
The experiments were conducted using the four-fold cross-validation setup introduced in the TUT Acoustic Scenes 2016 development and 2017 development datasets.

As the input of each system, the 64-dimensional log mel-band energy, which was calculated for each 40 ms time frame with 50\% overlap, was used.
The adjacency matrix ${\bf A}$ was calculated by counting the number of co-occurring sound events in each sound clip over the training dataset and normalizing the result in the range from 0 to 1.
After obtaining the output ${\bf y}_{t}$, active sound events were predicted by thresholding using an adaptive thresholding technique \cite{Xu_DCASE2017_01}.
The detection performance was evaluated by the F1-score and error rate in the segment-based metrics \cite{Mesaros_AS2016_01}, in which the segment length is set to 40 ms.
The other recording conditions and experimental conditions are listed in Table~\ref{tab:Condition}.
%
%
\begin{figure}[t]
\vspace{-1pt}
\centering
\hspace{-3pt}
\includegraphics[width=1.01\columnwidth]{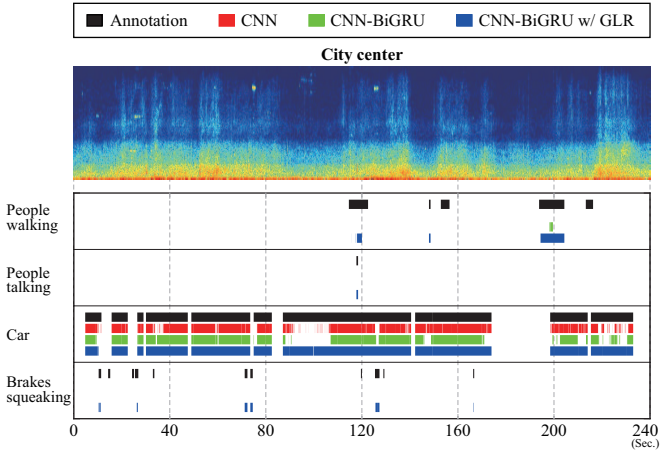}
\vspace{-22pt}
\caption{Annotations and event detection results of sounds recorded in city center. Only sound events occurring in the annotations are described.}
\label{fig:b098}
\end{figure}
%
%
\begin{figure}[t]
\centering
\hspace{-3pt}
\includegraphics[width=0.98\columnwidth]{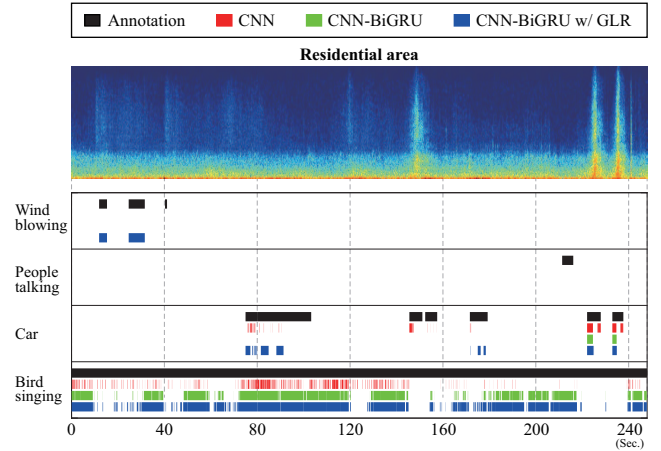}
\vspace{-10pt}
\caption{Annotations and event detection results of sounds recorded in residential area. Only sound events occurring in the annotations are described.}
\label{fig:a003}
\end{figure}
%
%
\subsection{Experimental results}
\label{subsec:results}
Table~\ref{tab:Result1} shows the F1-scores and error rates for CNN, CNN-BiGRU, and CNN-BiGRU with graph Laplacian regularization (GLR).
The results show that the proposed method moderately improves the SED performance in terms of both the F1-score and error rate.
In this experiment using the TUT Sound Events and Acoustic Scenes datasets, the proposed method improves the average SED performance by 7.9 percentage points from that of conventional CNN-BiGRU in terms of the F1-score.

To examine the detection results in more detail, we illustrate examples of annotations and the predicted results of sound events in Figs.~\ref{fig:b098} and \ref{fig:a003}.
The results also show that the proposed method detects sound events more accurately than the conventional methods.
Moreover, the results show that the proposed method can detect co-occurring sound events more accurately than the conventional methods.
For instance, the sound events ``car'' and ``brakes squeaking'' can be detected by adopting graph Laplacian regularization, whereas the conventional methods do not detect ``brakes squeaking'' events.
Thus, we conclude that graph Laplacian regularization based on the co-occurrence of sound events is a promising technique for SED.
%
\vspace{-6pt}
\section{Conclusion}
\label{sec:conclusion}
\vspace{-5pt}
In this paper, we proposed the neural-network-based SED with graph Laplacian regularization based on the co-occurrence of sound events.
Unlike conventional CNN or CNN-BiGRU-based SED methods, the proposed method can detect sound events with prior information on the co-occurrence of sound events.
This enables sound events to be modeled effectively and efficiently even if there are many types of sound events to model and limited training data.
The experimental results obtained using the TUT Sound Events 2016, 2017, and TUT Acoustic Scenes 2016 datasets show that the proposed method improves the SED performance by 7.9 percentage points in terms of the segment-based F1-score.
The experimental results also show that the proposed method can detect sound events that tend to co-occur, such as sound events ``car'' and ``brakes squeaking'', more accurately than the conventional methods.
%
%
\bibliographystyle{IEEEbib}
\bibliography{IEEEabrv,ICASSP2019refs,KeisukeImoto06}

\end{document}